 \documentclass[authoryear,preprint,review,12pt]{elsarticle}

\usepackage{amssymb}

\usepackage{url}

\usepackage{amsmath,amsfonts,amssymb,amsthm,mathrsfs}

\usepackage{graphicx}

\usepackage{algorithm}
\usepackage{algpseudocode}

\usepackage{bm}
\newcommand{\mathbold}[1]{\bm{#1}}
\newcommand{\mbf}[1]{\mathbf{#1}}

\newcommand{\T}{\mathsf{T}}    %
\newcommand{\N}{\mathcal{N}}   %

\DeclareMathOperator{\gammad}{Gamma}

\newcommand{\vtheta}[0]{\mathbold{\theta}}

\newcommand{\ve}{\mbf{e}}

\newcommand{\vm}{\mbf{m}}

\newcommand{\vu}{\mbf{u}}
\newcommand{\vv}{\mbf{v}}

\newcommand{\vx}{\mbf{x}}
\newcommand{\vy}{\mbf{y}}

\newcommand{\MA}{\mbf{A}}

\newcommand{\MH}{\mbf{H}}

\newcommand{\MK}{\mbf{K}}

\newcommand{\MP}{\mbf{P}}
\newcommand{\MQ}{\mbf{Q}}
\newcommand{\MR}{\mbf{R}}
\newcommand{\MS}{\mbf{S}}

\newcommand{\MW}{\mbf{W}}

\journal{Digital Signal Processing}

\begin{document}

\begin{frontmatter}

\title{Combining Particle MCMC with Rao-Blackwellized Monte Carlo Data Association for Parameter Estimation in Multiple Target Tracking}

\author{Juho Kokkala\corref{cor1}}
\ead{juho.kokkala@aalto.fi}
\ead[url]{http://becs.aalto.fi/~jkokkala}
\cortext[cor1]{Corresponding author}
\author{Simo S\"{a}rkk\"{a}\corref{cor2}}
\ead{simo.sarkka@aalto.fi}
\ead[url]{http://becs.aalto.fi/~ssarkka}

\address{Aalto University\\
Espoo, Finland\\}

\begin{abstract}
We consider state and parameter estimation in multiple target tracking problems with data association uncertainties and unknown number of targets. We show how the problem can be recast into a conditionally linear Gaussian state-space model with unknown parameters and present an algorithm for computationally efficient inference on the resulting model. The proposed algorithm is based on combining the Rao-Blackwellized Monte Carlo data association algorithm with particle Markov chain Monte Carlo algorithms to jointly estimate both parameters and data associations. Both particle marginal Metropolis--Hastings and particle Gibbs variants of particle MCMC are considered. We demonstrate the performance of the method both using simulated data and in a real-data case study of using multiple target tracking to estimate the brown bear population in Finland.
\end{abstract}


\end{frontmatter}

\section{Introduction}
\label{sec:introduction}
This paper is concerned with multiple target tracking (MTT), that is, with the problem of estimating the locations or states of several moving objects (targets) based on noisy measurements \citep[see, e.g., ][]{blackman:1999, bar-shalom:2001, kirubarajan:2005, mahlerbook:2007, challa:2011}. The challenge in MTT is that in addition to estimating the locations, one needs to solve the subproblems of estimating the number of targets and determining which target each measurement comes from, known as the data association problem. MTT methods have been applied, for example, to aircraft tracking \citep{hwang:2004}, video surveillance \citep{rao:2013}, evolutionary clustering \citep{mestre:2013}, and estimating the size of animal population \citep{abbas:2011}. In this paper we formulate the multiple target tracking problem as a Rao-Blackwellized particle filtering problem following \citet{sarkka:2007} and then show how we can use state-of-the-art particle Markov chain Monte Carlo (PMCMC) 
methods \citep{andrieu:2010} to estimate the parameters of the model.

In the Rao-Blackwellized Monte Carlo data association (RBMCDA) algorithm proposed by \citet{sarkka:2007}, target movements and measurements given targets are assumed to follow a linear-Gaussian state-space model. Thus, conditional on the data associations, posterior distributions for the target locations are obtained in closed form using the Kalman filter \citep{kalman:1960}. This enables the use of the Rao-Blackwellized particle filter \citep[RBPF, ][]{akashi:1977,doucet:2000-on,doucet:2000-rao,chen:2000,sarkka:2013} to sample the data associations. \citet{vihola:2007} proposed a similar RBPF filter, where the conditional linear-Gaussian model is formulated in the random set framework. 
In this paper we show how the RBMCDA algorithm of \citet{sarkka:2007} can be extended to joint estimation of unknown parameters along  with the target states. In the Bayesian framework \citep{gelman:2013,sarkka:2013}, parameters are modeled as random variables and the goal of parameter estimation is to compute the posterior probability distributions over parameters conditional on observations. Markov chain Monte Carlo (MCMC) methods are typically used to produce samples from the posterior distributions. In the context of state-space models, such as tracking problems, one needs to jointly sample both from the posterior of the parameters and the posterior of the states. Particle MCMC (PMCMC) algorithms \citep{andrieu:2010} are a special class of MCMC algorithms that use particle filter algorithms  to produce samples of state variables within MCMC. In this paper, we propose combining the RBMCDA and PMCMC algorithms to sample from the joint posterior distribution of data associations and parameters. This 
combined algorithm is intended for models where the movement and measurements from individual targets follow a linear-Gaussian state-space model conditional on the fixed number of unknown parameters. However, it is also possible to treat approximately linear-Gaussian state-space models by replacing the Kalman filters with extended Kalman filters (EKF), unscented Kalman filters (UKF), or other non-linear filters \citep[see, e.g.,][]{sarkka:2013}.

Using PMCMC in MTT has been suggested previously by \citet{vu:2014} and \citet{duckworth:2012}. These approaches use MCMC to propose data associations and the particle filter to sample target states conditional on the data associations. The algorithm of \citet{vu:2014} does not sample static parameters at all, while \citet{duckworth:2012} samples static parameters within the particle filter. Our proposed algorithm differs from these in that the MCMC is used to propose static parameters while data associations and the number of targets are sampled in the RBMCDA filter.

The use of PMCMC in combination with Rao-Blackwellized particle filters has been proposed before in other contexts \citep[e.g.][]{chopin:2010,peters:2010}. However, the particular method proposed in this article is novel since the combination of PMCMC and Rao-Blackwellized particle filters, in particular RBMCDA, has, to our knowledge, not been used in the multiple target tracking context.

The remainder of the article is structured as follows. In Section~\ref{sec:MTT}, we present a brief survey of the multiple target tracking literature. In Section~\ref{sec:pmcmc}, we review the particle filtering and particle MCMC algorithms, and in Section~\ref{sec:rbmcda} the RBMCDA algorithm. In Section~\ref{sec:pmcmc_rbmcda}, we present the combined RBMCDA--PMCMC algorithms. In the numeric experiments in Section~\ref{sec:pmcmc_rbmcda}, we use simulated data to compare the performance of the particle Gibbs with varying numbers of particles. We also present a real-data application of the algorithm to estimating the bear population of Finland based on a database of field-signs and direct observations. Pseudocodes for the algorithms are presented in \ref{app:algs}.

\subsection{Multiple Target Tracking Methods}
\label{sec:MTT}
Various filtering approaches for multiple target tracking have been proposed in literature. Joint probabilistic data association (JPDA) \citep{fortmann:1980} approximates the joint density by a Gaussian distribution. In the update step, the measurements are weighted by data association probabilities. In multiple hypothesis tracking (MHT) \citep{reid:1979,blackman:2004}, target state distributions are maintained for different data association histories. To prevent combinatorial explosion, heuristics are employed to discard unlikely hypotheses.

Multiple particle filtering \citep{bugallo:2007,djuric:2009} is based on tracking each target with a separate particle filter and approximatively combining the information in the weight update. More recently, \citet{closas:2012} proposed a refinement where the weight computation is iterated in a game-theory-inspired manner. Another approach based on partitioning the state is the Independent partition particle filter \citep{orton:2002}, where the state is partitioned so that states of clearly separate targets are sampled independently. \citet{yi:2013} provided a different view of the target independence approximation - they used the assumption to improve approximation of the filter predicted density rather than for independently propagating target states.

Random set based MTT approaches such as probability hypothesis density (PHD) \citep{mahler:2003,mahlerbook:2007} filtering are based on the theory of finite set statistics \citep[FISST, ][]{mahlerbook:2007}. The joint random set distribution is often approximated with the PHD, which a density whose integral gives the expected number of targets in the region. The PHD may further be approximated by Gaussian mixtures \citep{vo:2006} or particle filters \citep{vo:2003}. For the particle filter PHD approach, \citet{clark:2007-multi} proposed to assign the particles to target labels by expectation--maximization or k-means clustering. \citet{clark:2007-gaussian} proposed a particle PHD filter where the particles represent a mixture of Gaussians rather than point masses. In cardinalized PHD \citep{mahler:2007}, the probability distribution over number of targets is propagated along the PHD. Multi-target multi-Bernoulli filtering (MeMBer) \citep{vo:2009} is based on target-wise densities and independent 
existence probabilities. More recently, \citet{ravindra:2012} proposed a MeMBer filter where the independence of existence probabilities is preserved by modifying the posterior densities of targets while preserving the random finite set (RFS) density. A related idea is the set JPDA method \citep{svensson:2011} where the posterior after JPDA update is modified to improve Gaussian mixture estimation while preserving the RFS density. Recently, \citet{svensson:2014}  proposed formulating the multiple target tracking problem as computing posterior distributions over random finite sets of trajectories directly, rather than random finite sets of states. 

The RBMCDA \citep{sarkka:2007} algorithm used in this paper is based on assuming linear-Gaussian target dynamics and measurements and then using a Rao-Blackwellized particle filter, where Kalman filter is used to track target states and the particle approximation to approximate the distribution over data associations. A related idea by \citet{vihola:2007} proposed a RBPF filter, where the conditional linear-Gaussian model is formulated in the random set framework. \citet{petetin:2014} used a Rao-Blackwellized particle filter within the PHD framework.

\section{Particle Filtering and Particle MCMC}
\label{sec:pmcmc}
Consider a state-space model \citep[see, e.g., ][]{sarkka:2013} with measurements $\vy_1,\ldots,\vy_T \in \mathbb{R}^m$, hidden states $\vx_0,\ldots,\vx_T \in \mathbb{R}^n$, and parameters $\vtheta \in \mathbb{R}^d$, which consists of the Markovian dynamic model
\begin{equation}
  \vx_k \sim p(\vx_k \mid \vx_{k-1}, \vtheta)
\end{equation}
and the measurement model
\begin{equation}
 \vy_k \sim p(\vy_k \mid \vx_k, \vtheta).
\end{equation}
When the parameters $\vtheta$ are fixed, the state sequence $\vx_{0:T}$ is assumed to be Markovian and the measurements are assumed to be conditionally independent given the states. In the following, we briefly review the particle filtering (sequential importance resampling, SIR) algorithm for approximating the filtering distributions of the states, that is, $p(\vx_k \mid \vy_{1:k}, \vtheta)$ and the particle MCMC algorithms that combine particle filtering with MCMC to sample from the joint posterior of the parameters and the states, $p(\vtheta,\vx_{0:T} \mid \vy_{1:T})$. 

\subsection{Particle filtering}
\label{sec:pf}

In sequential importance resampling type particle filtering \citep{doucet:2000-on}, the filtering distribution at time step $k$, $p(\vx_k \mid \vy_{1:k})$, is approximated by a finite set of $N$ discrete particles with weights, $\{(w_k^{(i)},\tilde{\vx}_k^{(i)}) : i=1,\ldots,N\}$. This is interpreted as the density approximation 
\begin{equation}
 p(\vx_k \mid \vy_{1:k} ) \approx \sum_{i=1}^N w_k^{(i)}\delta(\vx_k - \tilde{\vx}^{(i)}_k),
\end{equation}
where $\delta(\cdot)$ is the Dirac delta function. The particle filtering algorithm iterates the following steps through the measurements $k=1,\ldots,T$:
\begin{enumerate}
 \item Sample new particles from an importance distribution: $\tilde{\vx}_{k}^{(i)} \sim \pi(\vx_{k} \mid \tilde{\vx}_{k-1}^{(i)}, \vy_{k})$.
 \item Compute updated weights: $v^{(i)}_k = w^{(i)}_{k-1}\,\frac{p(\tilde{\vx}_{k}^{(i)})\,p(\vy_k \mid \tilde{\vx}_{k}^{(i)})}{\pi(\vx_{k} \mid \tilde{\vx}_{k-1}^{(i)}, \vy_{k})}$.
 \item Normalize weights: $w^{(i)}_k = \frac{v^{(i)}_k}{\sum_i v^{(i)}_k }$.
 \item Resample: if necessary, draw $N$ new particle values $\tilde{\vx}_{k}^{(i)}$ from the original $\tilde{\vx}_{k}^{(i)}$ with probabilities $w_k$.
\end{enumerate}

The purpose of the resampling step is to avoid degeneracy where one particle attains all weight. It may be performed periodically with a fixed interval or adaptively based on effective sample size \citep{liu:1995} declining below a threshold. 

For purposes of parameter estimation, the particle filter can also be used to form an approximation to the marginal likelihood $p(\vy_{1:T} \mid \vtheta)$ \citep[see, e.g.,][]{andrieu:2004,sarkka:2013}:
\begin{equation} 
\hat{p}(\vy_{1:T} \mid \vtheta) = \prod_{k=1}^T\hat{p}(\vy_k \mid \vy_{1:k-1},\vtheta), \label{eq:pflikelihood}
\end{equation}
where 
\begin{equation}
 \hat{p}(\vy_k \mid \vy_{1:k-1},\vtheta) = \sum_{i=1}^N v^{(i)}_k. 
\end{equation}
When combined with Markov chain Monte Carlo (MCMC), this leads to so called particle MCMC (PMCMC) methods \citep{andrieu:2010}.

\subsection{Rao-Blackwellized Particle Filter}
For models, where the filtering problem is analytically tractable conditional on some subset of variables, one may reduce the variance of the importance weights by the Rao-Blackwellized particle filter \citep{akashi:1977,doucet:2000-on,doucet:2000-rao,chen:2000}, where the particle filter is employed only for the non-analytically tractable subset, and the tractable part is marginalized analytically. For example, in conditionally linear-Gaussian models of the form
\begin{equation}
\begin{split}
\vx_k  &\sim \N(\MA_{k-1}(\vu_{k-1}) \, \vx_{k-1}, \MQ_{k-1}(\vu_{k-1})) \\
\vy_k  &\sim \N(\MH_k(\vu_k) \, \vx_k,\MR_k(\vu_k))   \\
\vu_k  &\sim p(\vu_k \mid \vu_{k-1}),
\end{split}
\label{eq:CLGSS}
\end{equation}
the particles of the Rao-Blackwellized particle filter contain samples of the latent variables $\vu_k$, and the states $\vx_k$ are marginalized out using the Kalman filter. Although we usually assume that the latent variables are a priori Markovian, the algorithm generalizes without modification to the non-Markovian (but causal) case. That is, the last equation above may be generalized to $p(\vu_k \mid \vu_{1:k-1})$.

\subsection{Particle MCMC}
The idea of using particle filters within a Markov chain Monte Carlo (MCMC) sampler was suggested by, for example, \citet{fernandez-villaverde:2007,jones:2010}. Theoretical justification that these particle MCMC algorithms indeed produce Markov chains that converge to the joint posterior of the states and parameters was provided by \citet{andrieu:2010}. In this section, we discuss two different particle MCMC algorithms, both introduced by \citet{andrieu:2010}. First, we discuss particle marginal Metropolis--Hastings, which is based on the likelihood approximation produced by the particle filter. Second, we discuss particle Gibbs where a modification of the particle filter called conditional sequential Monte Carlo is used to move in the space of state sequences. 

The particle marginal Metropolis--Hastings (PMMH) algorithm is a variant of the Metropolis--Hastings algorithm, where the exact evaluation of the likelihood (and posterior) is replaced by running the particle filter and using the approximate likelihood. The algorithm is initialized by selecting initial parameters $\vtheta^0$ and running the particle filter to obtain approximate marginal likelihood $\hat{p}(\vtheta \mid \vy_{1:T})$. Then, the algorithm produces samples from the parameters and particle sets, $(\vtheta^1,\vx^{1,(1:N)}_{1:T},w^{1,(1:N)}_T),(\vtheta^2,\vx^{2,(1:N)}_{1:T},w^{2,(1:N)}_T),\ldots$ by iterating the following steps
\begin{enumerate}
\item Draw proposed parameters: $\vtheta^\ast \sim q(\vtheta^\ast \mid \vtheta^{j-1})$
\item Run the particle filter (Section~\ref{sec:pf}) using the parameters $\vtheta^\ast$ to obtain weighted set of particles $(w^{\ast,(1:N)}_T,\vx^{\ast,(1:N)}_{1:T})$ and a marginal likelihood estimate $\hat{p}(\vy_{1:T} \mid \vtheta^\ast)$ (Eq.~\ref{eq:pflikelihood})
\item With probability 
\begin{equation}
 \alpha_j = \min\left(1, \frac{q(\vtheta^{j-1}\mid\vtheta^\ast)}{q(\vtheta^\ast \mid \vtheta^{j-1})}\, \frac{\hat{p}(\vy_{1:T} \mid \vtheta^\ast)}{\hat{p}(\vy_{1:T} \mid \vtheta^{j-1})} \, \frac{p(\vtheta^\ast)}{p(\vtheta^{j-1})}    \right)
\end{equation}
accept the proposal, that is:
\begin{align}
& \left(\vtheta^j,~w^{j,(1:N)}_T,~\vx^{j,(1:N)}_{1:T},~\hat{p}(\vy_{1:T} \mid \vtheta^j) \right) \nonumber \\
:= & \left(\vtheta^\ast,~w^{\ast,(1:N)}_T,~\vx^{\ast,(1:N)}_{1:T},~\hat{p}(\vy_{1:T} \mid \vtheta^\ast) \right).
\end{align}
\item If the proposal is not accepted, copy the values from previous iteration:
\begin{align}
& \left(\vtheta^j,~w^{j,(1:N)}_T,~\vx^{j,(1:N)}_{1:T},~\hat{p}(\vy_{1:T} \mid \vtheta^j) \right) \nonumber \\
:= & \left(\vtheta^{j-1},~w^{j-1,(1:N)}_T,~\vx^{j-1,(1:N)}_{1:T},~\hat{p}(\vy_{1:T} \mid \vtheta^{j-1}) \right).
\end{align}
\end{enumerate}

Samples from the state, $\vx^j_{1:T}$, may be obtained by drawing one particle from the accepted particles $\vx^{j,(1:N)}_{1:T}$ with using the importance weights $w^{j,(1:N)}_T$ as probabilities. The Markov chain produced by the PMMH algorithm is ergodic in an extended space consisting of the parameters and the particle sets so that the marginal stationary distribution in the states-and-parameters space is the correct posterior distribution \citep{andrieu:2010}. The particle Metropolis-Hastings algorithm may also be interpreted as a Multiple Try Metropolis algorithm, as \citet{Martino:2015} point out.

The particle Gibbs algorithm is an MCMC algorithm moving in the joint space of $(\vtheta,\vx_{1:T})$. The particle Gibbs uses a regular MCMC, namely Gibbs sampling, step to draw new parameter values conditional on the states and a variant of particle filter, conditional SMC, to sample new states. The conditional SMC is a variant of the particle filter that takes the current state sequence as input and fixes the states for one particle to the input sequence instead of drawing them from the importance distributions. That is, instead of drawing $\vx^{(1)}_k$ from the importance distribution $q(\vx_k \mid \vx^{(1)}_{k-1},\vy_k)$, the value of $\vx^{(1)}_k$ is set to the old value of $\vx_k$. For particles $2,\ldots,N$ the algorithm proceeds exactly as the particle filter. Note that the weights are nevertheless recomputed even for the fixed particle as if the states were sampled from the importance distribution. After running the CSMC, $\vx^j_{1:T}$ is sampled among the particles using the 
importance weights. In total, the particle Gibbs algorithm iterates the following steps:
\begin{enumerate}
\item Draw $\vtheta^j \sim p(\vtheta \mid \vx^{j-1}_{1:T})$
\item Generate $(\vx^{j,(1:N)}_{1:T},w^{j,(1:N)}_{1:T})$ by running the conditional SMC using parameters $\vtheta^j$ and fixing the first particle to $\vx^{j-1}_{1:T}$.
\item Draw $\vx^j_{1:T}$ from $\vx^{j,(1:N)}_{1:T}$ with probabilities $w^{j,(1:N)}_T$.
\end{enumerate}

Since the joint posterior distribution $p(\vx_{0:T},\vtheta \mid \vy_{1:T})$ is an invariant distribution for both the CSMC move and the parameter sampling move, the resulting particle Gibbs algorithm is a MCMC sampler targeting the joint posterior distribution \citep{andrieu:2010}.

\citet{andrieu:2010} also show that it is possible to improve the MCMC estimates by using the state sequences produced by all particles rather than only one state sequence selected per MCMC step. In particle Gibbs, all particles may  be taken as samples weighted by their respective importance weights. Furthermore, in PMMH one may also use the particles corresponding to rejected parameter proposals by weighting the new particle set and the particle set corresponding to the last accepted proposal by the Metropolis--Hastings acceptance probability. 

Combining Rao-Blackwellized particle filters with PMCMC was already suggested by \citet{chopin:2010} and \citet{peters:2010}. Naturally, since the RBPF is a particle filter in the state space of the latent variables $\vu$, using it in a PMCMC algorithm produces a MCMC sampler targeting the joint posterior $p(\vu_{0:T},\vtheta \mid \vy_{1:T})$. \citet{whiteley:2010} combined the discrete particle filter \citep{fearnhead:2003} with PMCMC to do inference in switching state-space models. In addition, Rao-Blackwellized PMCMC has been used by  \citet{nevat:2011} in channel tracking in wireless relay networks, by \citet{minvielle:2014} in an electromagnetic inverse problem, and by \citet{peters:2013} in the context of a financial commodity model.

\section{Rao-Blackwellized Monte Carlo Data Association}
\label{sec:rbmcda}
In this section, we review the RBMCDA algorithm proposed by \citet{sarkka:2007}. The algorithm is formulated for models where the target dynamics are linear with Gaussian process noise, and the measurements conditional on data associations are a linear function of target states plus Gaussian measurement noise. However, as was shown in \citet{sarkka:2007}, it is also possible to handle non-linear state-space models by replacing the Kalman filters in the algorithm non-linear extensions such as extended Kalman filters (EKF), unscented Kalman filters (UKF), or more general non-linear Gaussian filters \cite{sarkka:2013}.

We denote the state of the $j$th target at $k$th time step by $\mathbf{x}_{k,j}$. The dynamics are assumed to be linear with Gaussian noise, that is,
\begin{equation}
p(\vx_{k,j} \mid \vx_{k-1,j}) = \N(\vx_{k,j} \mid \MA_{k-1}\vx_{k-1,j},\MQ_{k-1}),
\end{equation}
where $\MA_{k-1}$ is the time dependent transition matrix and $\MQ_{k-1}$ is the time dependent process noise covariance matrix. The dynamics of different targets are assumed to be independent. The measurement model is such that each measurement corresponds to a randomly selected target, denoted by $c_k$ and conditional on the association, the measurement depends only on the state of target $c_k$. In particular, the measurements conditional on target states and associations are linear Gaussian:
\begin{equation}
p(\vy_k \mid \vx_{k,j}, c_k = j) = \N(\vy_k \mid \MH_{k}\vx_{k,j},\MR_{k}),
\end{equation}
where $\MH_{k}$ is the measurement matrix and $\MR_{k}$ is the measurement noise covariance matrix. 

Unknown and varying number of targets is handled by defining an indicator variable $\ve_k$ which tells which of the targets are alive at the current time step. The initial state has no targets and the targets are assumed to enter the state at the time of their first observation. Targets are removed from consideration by setting the indicator to $0$ after a target has not been observed for a while. \citet{sarkka:2007} also consider removing targets probabilistically based on time since last observation. Since the targets are labeled according to the order they are first observed, the data association prior $p(c_k \mid c_{k-1},\ldots,c_1,\ve_{k-1})$ contains positive probabilities only for the targets contained in $c_1,\ldots,c_{k-1}$ that are visible in $\ve_{k-1}$ as well as one new target. Clutter measurements, that is, measurements that are not related to any 
target, are modeled by specifying that $p(\vy_k \mid c_k = 0, \vx_{k,:})$ is some fixed distribution independent of the target states.  The state of a new target at the time of its first observation is assumed to follow $\N(\vm_0,\MP_0)$. The resulting RBMCDA filter is shown in pseudocode in Algorithm~\ref{alg:rbmcda}.

The model defined above is of the conditionally linear-Gaussian form \eqref{eq:CLGSS} so that the latent variable $\vu_k$ consists of the data association $c_k$ and the visibility indicator $\ve_{k}$. Thus, a RBPF may be applied. Furthermore, since the state-space of possible data associations is finite, the optimal importance distribution may be used for sampling the data association $c_k$.

In practice, a computational speedup may be obtained by performing the Kalman filter prediction and updates need only for each unique data association history instead of all particles, some of which are identical. For simplicity of the presentation, this speedup is not explicitly written out in Algorithm~\ref{alg:rbmcda}.

The algorithm state consists of $N$ particles that represent an approximation of the posterior distribution over data association histories at step $k$. The following information is stored for each particle $i\in\{1,\ldots,N\}$:
\begin{equation}
\textrm{Particle}_i = \left(c^{(i)}_{1:k}, \vm_{k,1}^{(i)},\vm_{k,2}^{(i)},\ldots,\vm_{k,T_k^{(i)}},\MP_{k,1}^{(i)},\MP_{k,2}^{(i)},\ldots,\MP_{k,T_k^{(i)}}, w^{(i)}_k \right),
\end{equation}
where
\begin{itemize}
 \item $c^{(i)}_{1:k}$ is the data association history for measurements $1,\ldots,k$ 
\item $T_k^{(i)}$ is the number of different targets seen so far, i.e., maximum of $c^{(i)}_{1:k}$ 
\item $\vm_{k,j}^{(i)},\MP_{k,j}^{(i)}$ are the mean and covariance of the distribution of the state of target $j$ conditional on $c^{(i)}_{1:k}$
 \item $w_k^{(i)}$ is the importance weight of the particle.
\end{itemize}
The algorithm proceeds through the measurements as follows:
\begin{enumerate}
\item For all particles $i \in \{1,\ldots,N\}$:
\begin{enumerate}
\item For all targets $j\in \{1,\ldots,T^{(i)}_k\}$, propagate the target state distribution moments through the Kalman filter prediction step to obtain the moments $\vm^{(i)-}_{k,j},\MP^{(i)-}_{k,j}$ of the distributions $p(\vx_{k,j} \mid \vy_{1:k-1}, c^{(i)}_{1:k-1})$.
\begin{equation}
 \vm^{(i)-}_{k,j} = \MA_{k-1}\,\vm^{(i)}_{k-1,j},~\MP^{(i)-}_{k,j} = \MA_{k-1}\,\MP_{k-1,j}\,\MA^\T_ {k-1} + \MQ_{k-1}.
\end{equation}
\item For all targets $j\in \{1,\ldots,T^{(i)}_k\}$ , run the Kalman filter update step conditional on the data association to obtain the moments $\vm_{k,j}^{(i)\ast}, \MP_{k,j}^ {(i)\ast}$ of the distributions $p(\vx_{k,j} \mid \vy_{1:k}, c_{1:k-1}^{(i)}, c_k = j)$ and the likelihoods $p(\vy_k \mid c^{(i)}_{1:k}, \vy_{1:k-1})$ (See Algorithm~\ref{alg:update}).
\item Evaluate the optimal importance distribution $$P(c_k = j) = \frac{p(c_k = j \mid c^{(i)}_{1:k})\,p(\vy_k \mid c^{(i)}_{1:k}, \vy_{1:k-1})}{\sum_{j=1}^{T^{(i)}_k+1}p(c_k = j \mid c^{(i)}_{1:k})\,p(\vy_k \mid c^{(i)}_{1:k}, \vy_{1:k-1})}$$
\item Draw $c^{(i)}_k$ from the optimal importance distribution
\item Set $(\vm^{(i)}_{k,c_k},\MP^{(i)}_{k,c_k}) = (\vm^{(i)\ast}_{k,c_k},\MP^{(i)\ast}_{k,c_k})$, 
\item For $j \neq c_k$: set $(\vm^{(i)}_{k,j},\MP^{(i)}_{k,j}) = (\vm^{(i)-}_{k,j},\MP^{(i)-}_{k,j})$, the predicted distributions
\item Update particle weight: $w^{(i)}_k := w^{(i)}_{k-1}\,\sum_{j=1}^{T^{(i)}_k+1}p(c_k = j \mid c^{(i)}_{1:k})\,p(\vy_k \mid c^{(i)}_{1:k}, \vy_{1:k-1})$
\end{enumerate}
\item Normalize particle weights to sum to unity
\item Possible resampling step
\end{enumerate}
The marginal likelihood approximation similar to Eq.~\ref{eq:pflikelihood} in Section~\ref{sec:pf} is computed by 
\begin{equation} 
\hat{p}(\vy_{1:T} \mid \vtheta) = \prod_{k=1}^T\hat{p}(\vy_k \mid \vy_{1:k-1},\vtheta), 
\end{equation}
where 
\begin{equation}
 \hat{p}(\vy_k \mid \vy_{1:k-1},\vtheta) = \sum_{i=1}^Nk w^{(i)}_{k-1} \left(\sum_{j=1}^{T^{(i)}_k+1}p(c_k = j \mid c^{(i)}_{1:k})\,p(\vy_k \mid c^{(i)}_{1:k}, \vy_{1:k-1}) \right).
\end{equation}

\section{PMCMC for RBMCDA}
\label{sec:pmcmc_rbmcda}
In this section, we show how the RBMDCA algorithm described in Section~\ref{sec:rbmcda} can be combined with the PMCMC algorithms described in Section~\ref{sec:pmcmc}. The model is assumed to be of the linear-Gaussian form specified in Section~\ref{sec:rbmcda} with the extension that the dynamic model transition matrices $\MA_k(\vtheta)$, process noise covariances $\MQ_k(\vtheta)$, measurement model matrices $\MH_k(\vtheta)$ and measurement noise covariance matrices $\MR_k(\vtheta)$ depend on some parameter vector $\vtheta$ of fixed dimension.
The particle marginal Metropolis--Hastings is based on using the particle filter based likelihood approximation. In the RBMCDA context,  the PMMH algorithm produces a Markov chain moving in the joint space of the parameters and particle sets of the data associations, that is, the samples are of the form $(\vtheta^j,c^{j,(1:N)}_{1:T},w^{j,(1:N)}_T)$. The algorithm iterates the following at steps $j=1,2,\ldots$:
\begin{enumerate}
\item Draw proposed parameters: $\vtheta^\ast \sim q(\vtheta^\ast \mid \vtheta^{j-1})$
\item Run the RBMCDA filter (see Section~\ref{sec:rbmcda} or Algorithm~\ref{alg:rbmcda}) using the parameters $\vtheta^\ast$ to obtain weighted set of particles $(w^{\ast,(1:N)}_T,c^{\ast,(1:N)}_{1:T})$ and a marginal likelihood estimate $\hat{p}(\vy_{1:T} \mid \vtheta^\ast)$
\item With probability 
\begin{equation}
 \alpha_j = \min\left(1, \frac{q(\vtheta^{j-1}\mid\vtheta^\ast)}{q(\vtheta^\ast \mid \vtheta^{j-1})}\, \frac{\hat{p}(\vy_{1:T} \mid \vtheta^\ast)}{\hat{p}(\vy_{1:T} \mid \vtheta^{j-1})} \, \frac{p(\vtheta^\ast)}{p(\vtheta^{j-1})}    \right)
\end{equation}
accept the proposal, that is:
\begin{align}
& \left(\vtheta^j,~w^{j,(1:N)}_T,~c^{j,(1:N)}_{1:T},~\hat{p}(\vy_{1:T} \mid \vtheta^j) \right) \nonumber \\
:= & \left(\vtheta^\ast,~w^{\ast,(1:N)}_T,~c^{\ast,(1:N)}_{1:T},~\hat{p}(\vy_{1:T} \mid \vtheta^\ast) \right).
\end{align}
\item If the proposal is not accepted, copy the values from previous iteration:
\begin{align}
& \left(\vtheta^j,~w^{j,(1:N)}_T,~c^{j,(1:N)}_{1:T},~\hat{p}(\vy_{1:T} \mid \vtheta^j) \right) \nonumber \\
:= & \left(\vtheta^{j-1},~w^{j-1,(1:N)}_T,~c^{j-1,(1:N)}_{1:T},~\hat{p}(\vy_{1:T} \mid \vtheta^{j-1}) \right).
\end{align}
\end{enumerate}
Samples from the posterior of data associations are obtained by drawing from $(c^{j,(1)}_{1:T},c^{j,(2)}_{1:T},\ldots,c^{j,(N)}_{1:T})$ with probabilities $w^{j,(1:N)}_T$. 
In this work, we use symmetric multivariate Gaussian random-walk proposals for parameters. The covariance of the proposal distribution is adapted using the sample covariance of the samples produced so far, following the idea of \citet{haario:2001}. We adapt the covariance only during initial warmup to ensure the ergodicity of the adapting process is maintained in particle MCMC. The resulting RBMCDA--PMMH algorithm is shown in pseudocode in Algorithm~\ref{alg:rbmcda_pmmh}. 

Following the idea of conditional SMC, also the RBMCDA algorithm can be modified so that one particle is fixed to a given data association history. This also results in a MCMC move whose invariant distribution is the conditional posterior of data associations given parameters. This conditional RBMCDA algorithm is shown in pseudocode in Algorithm~\ref{alg:crbmcda}. Since in general models, the conditional posterior of parameters conditional on the data associations may not be available in closed-form, we replace the Gibbs step of PGibbs by Metropolis--Hastings steps for parameters. Thus, RBMCDA--PGibbs algorithm iterates the following steps:
\begin{enumerate}
\item Propose new $\vtheta^\ast \sim q(\vtheta^\ast \mid \vtheta^{j-1})$
\item With probability 
\begin{equation}
 \alpha_j = \min\left(1, \frac{q(\vtheta^{j-1}\mid\vtheta^\ast)}{q(\vtheta^\ast \mid \vtheta^{j-1})}\, \frac{p(\vy_{1:T} \mid \vtheta^\ast, c^{j-1}_{1:T})}{p(\vy_{1:T} \mid \vtheta^{j-1}, c^{j-1}_{1:T})} \, \frac{p(\vtheta^\ast)}{p(\vtheta^{j-1})}    \right),
\end{equation} accept the proposal, that is, set $\vtheta^j := \vtheta^\ast$. Else, set $\vtheta^j := \vtheta^{j-1}$.
\item Generate $(c^{j,(1:N)}_{1:T},w^{j,(1:N)}_{1:T})$ by running the conditional RBMCDA (Algorithm~\ref{alg:crbmcda} using parameters $\vtheta^j$ and fixing the data associations in the first particle to $c^{j-1}_{1:T}$.
\item Sample a data association sequence $c^j_{1:T}$ from $c^{j,(1:N)}_{1:T}$ with probabilities $w^{j,(1:N)}_T$.
\end{enumerate} 

To evaluate the acceptance ratios, the likelihood conditional on data associations, $p(\vy_{1:T} \mid \vtheta, c^{j-1}_{1:T})$, needs to be evaluated using the Kalman filter as shown in Algorithm~\ref{alg:likelihood}. For the Metropolis--Hastings proposal distributions $q$, we use the multivariate Gaussian random walk proposal adapted similarly as in the RBMCDA--PMMH algorithm. The resulting RBMCDA--PGibbs algorithm is shown in pseudocode in Algorithm~\ref{alg:rbmcda_pgibbs}. 

In some preliminary experiments, we observed that the conditional RBMCDA move sometimes led to poor mixing as the targets associated to early measurements usually did not change. To improve mixing, we also introduced additional Gibbs sampling steps where the targets associated to some particular measurements are redrawn from their conditional distributions. 

\section{Experimental Results}
\label{sec:numeric}
\subsection{Simulated Data}
In this section we compare the performance of the RBMCDA--PGibbs algorithms with varying number of particles. We generate a simulated dataset and run different MCMC algorithms to estimate the posterior distribution of parameters and data associations. We look at the convergence of the distribution of the number of targets in terms of Kolmogorov distance to a distribution obtained by a longer RBMCDA--PGibbs run. The Kolmogorov distance is compared against the total number of Kalman filter predict and update function calls. 

\begin{figure}[h]
\centering
\includegraphics{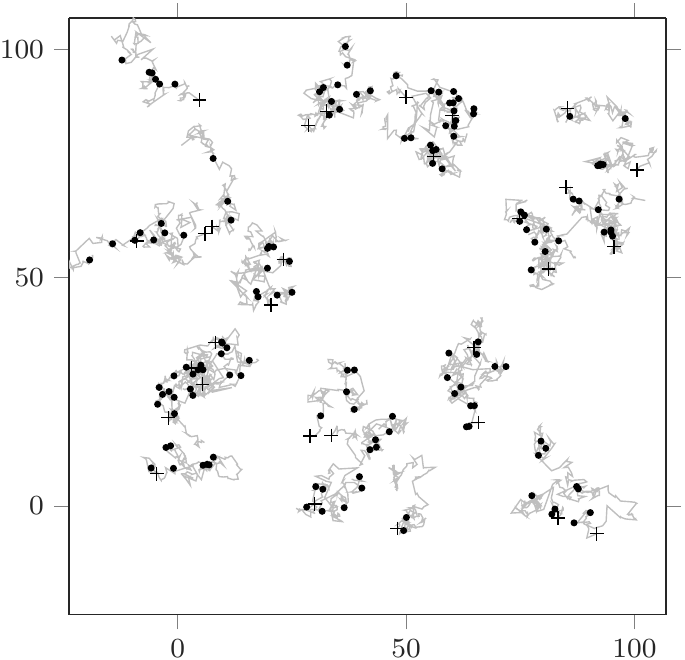}
\caption{Visualization of the simulated scenario. Trajectories of the targets are shown as gray lines, measurements as black dots and final target locations as black pluses.}
\label{fig:trajectories}
\end{figure}

We simulated $30$ two-dimensional target trajectories using the Ornstein--Uhlenbeck mean-reverting model:
\begin{equation}
\mathrm{d}\vx =  \lambda(\vx_0 - \vx)\mathrm{d}t + \sqrt{q}\mathrm{d}\MW,
\end{equation}
where $\vx$ is the target location and $\vx_0$ is a fixed mean location of the target. The parameters were set to $\lambda=0.5,\sqrt{q}=10$. The mean locations were sampled uniformly randomly in the window $[0,100]\times[0,100]$. Initial target locations were drawn from the steady-state distribution of the Ornstein--Uhlenbeck process. Then, $150$ observation times were sampled uniformly randomly in $[0,1]$, and data associations were generated so that the target associated to each measurement was selected randomly, but the data-associations were resampled until an association history where every target is obtained at least once was obtained. The measurements were the locations plus uncorrelated Gaussian noise with standard deviation $\sigma=0.5$ in both coordinates. The simulated target movements and observations are shown in Figure~\ref{fig:trajectories}.

10 chains of RBMCDA--PGibbs were run for $10^6$ steps. First half was discarded as warmup and the remaining samples from all chains pooled. The Ornstein--Uhlenbeck model was used for the target dynamics so that the target state is  $4$-dimensional consisting of the (constant) mean location and the actual location.  The initial density of new targets, $\N(\vm_0,\MP_0)$, was obtained by taking the sample mean and sample covariance of all observations, and using these for the distribution of the mean location coordinates. For the actual location coordinates, the corresponding steady-state distribution was used. Note that the initial density thus depends on the model parameters. For the model parameters we used Gamma priors with scale $2$ \citep{chung:2013} and modes ($\sqrt{q}=15,\lambda=1/3,\sigma=0.75$). These modes were selected so that the prior mode is somewhat off from the ground truth and favors a smaller number of targets. The data association prior $p(c_k \mid c_{1:k-1})$ was obtained as follows. The probability of new target is set to the conditional probability of a new target appearing conditional on a latent number of targets drawn uniformly from $\{1,\ldots,\textrm{number of observations}\}$ and each association being drawn uniformly from the latent number of targets. All old targets have equal probability. No clutter measurements nor target deaths were used.

\begin{figure}[h]
\centering
\includegraphics{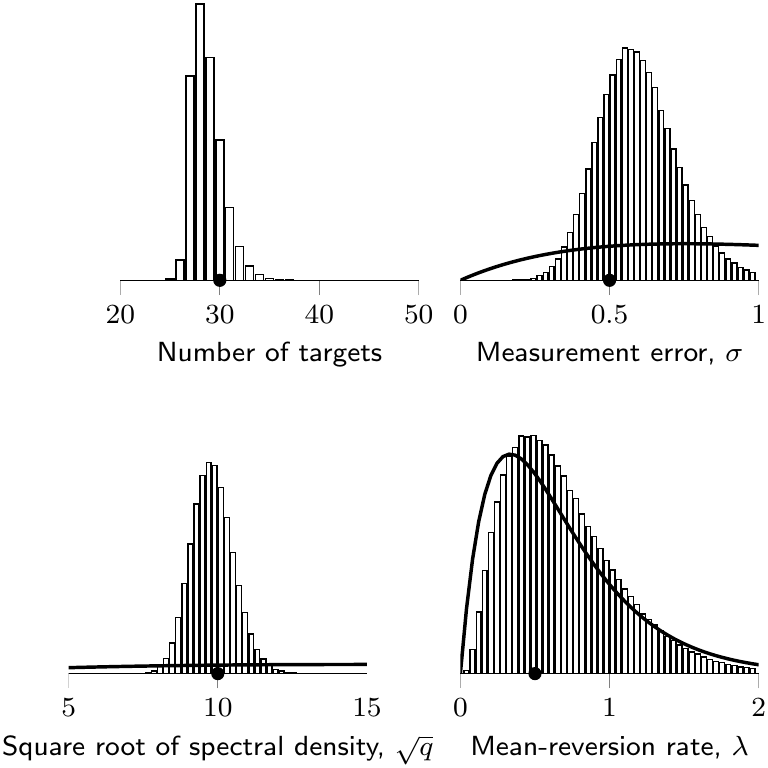}
\caption{Posterior distributions of the parameters ($\sqrt{q},\sigma,\lambda$) and the number of targets in the simulated scenario. The corresponding prior densities for parameters are shown as solid lines. Ground-truth parameters are marked as dots on the axis.}
\label{fig:histogram_simulated}
\end{figure}

We checked MCMC convergence using the potential scale reduction factor \citep{gelman:2013} with the implementation in GPStuff \citep{vanhatalo:2013}. Using the latter halves of the $10$ chains, the PSRF for all $3$ static parameters was below $1.01$, so we conclude that the chains have converged and pooling samples from the different chains is justified. The results are shown in Figure~\ref{fig:histogram_simulated}. The number of targets is slightly underestimated, which is natural as the parameter prior modes was set to favor a smaller number of targets compared to the true parameters. The posteriors of $\sqrt{q}$ and the measurement error $\sigma$ are clearly thinner than the prior and the modes are moved towards the truth. The posterior of the mean-reversion rate $\lambda$ is rather wide. This is explained by the fact that the time window of the simulation was quite short relative to the value of $\lambda$. However, the posterior of $\lambda$, too, was slightly moved towards true value.

To investigate the usefulness of parameter estimation, we also ran the RBMCDA--PGibbs with the same number of particles and chain lengths without sampling for parameters, that is, using the initial parameter values. We compared the accuracy based on the probability of the true number of targets as well as the OSPA metric \citep{schuhmacher:2008} for the posterior mean locations for all targets at the time of the $150$th measurement. The results are shown in Table~\ref{tab:ospa}. To save computational resources, the OSPA metric was computed using only every $500$th step of the MCMC chains.

\begin{table}
\caption{The simulated experiment. Comparison of RBMCDA--PGibbs with and without parameter estimation. Posterior probability of $30$ targets (the ground truth) as well as mean OSPA metric of the final target locations.}
\label{tab:ospa}
\begin{tabular}{c|cc}
Parameter estimation & P(Correct number of targets)        & Mean OSPA \\ \hline
Yes                  & $0.14$                              &   $5.95$         \\
No                   & $0.005$                             &  $8.02$     \\
\end{tabular}
\end{table}

\begin{figure}[h]
\centering
\includegraphics{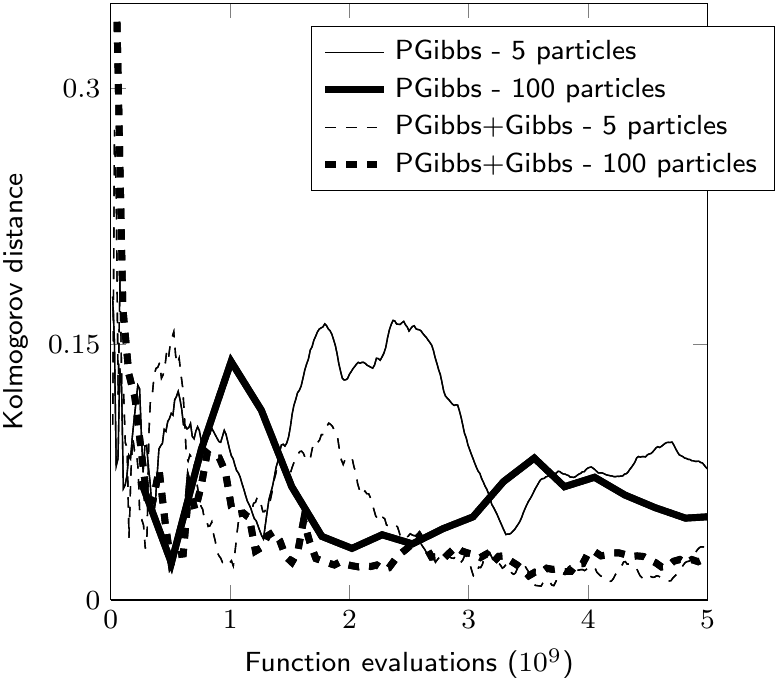}
\caption{Convergence of the distribution of the number of targets with varying algorithms. Kolmogorov distance vs. number of Kalman filter function evaluations. }
\label{fig:kolmogorov}
\end{figure}

We tried RBMCDA--PGibbs with $5$ and $100$ particles both with and without additional Gibbs steps. For each algorithm, 5 independent chains were used. Figure \ref{fig:kolmogorov} shows Kolmogorov distances to the distribution of Figure~\ref{fig:histogram_simulated} as a function of Kalman filter function evaluations. These are evaluated by cutting the chains at selected sample sizes, pooling results from all $5$ chains and discarding first half as warmup\footnote{To save computation time, the chains used for RBMCDA--PGibbs with Gibbs steps and  $5$ particles are $5$ first chains of the $10$ that were used to produce the ground truth. However, all samples used to this plot are discarded as warmup in the gold-standard distribution, so this is unlikely to bias the results.}.

\subsection{Real Data: Estimating Brown Bear Population}
\label{sec:case}
We consider a dataset of location records of direct sightings and field-sign observations of brown bears in Finland provided by \emph{Finnish Game and Fisheries Research institute}. The main quantity of interest in this study is the number of distinct packs (families) observed, which can then be used to estimate the overall population size by using an extrapolation factor \citep{kojola:2007}. We use a probabilistic approach for estimating the number of distinct families by formulating the problem as a multiple target tracking problem, where the targets are the packs. The posterior distribution for the number of packs is then obtained as a byproduct of the multiple target tracking solution. We used data of observations from year 2013 selecting only observations where cubs were present. We selected observations from one game management district (Kaakkois-Suomen riistanhoitopiiri).   \citet{abbas:2011} used RBMCDA in his Master's thesis for population estimation with this type of data, but this work did not use PMCMC for parameter estimation.

For target movement, we used the Ornstein--Uhlenbeck mean-reverting model (cf. Section~\ref{sec:numeric}), and measurement locations were assumed to be the actual target location plus Gaussian noise independent in both coordinates. Conditional on the parameters, the target dynamics of each year was assumed to be independent. Weakly informative $\gammad(2,\mu)$-priors were used for the parameters with modes: $\sqrt{q} = 2500~\mathrm{m}/\mathrm{d},~\lambda = 0.5~ \mathrm{d}^{-1},~\sigma = 100~\mathrm{m}$. We used $5$ particles and  $10$ separate MCMC chains were run for $100,000$ steps each. The results presented here are based on discarding 
the first half of each chain as warmup and combining the remaining samples from all  $10$ chains. 

\begin{figure}[h!]
\centering
\includegraphics{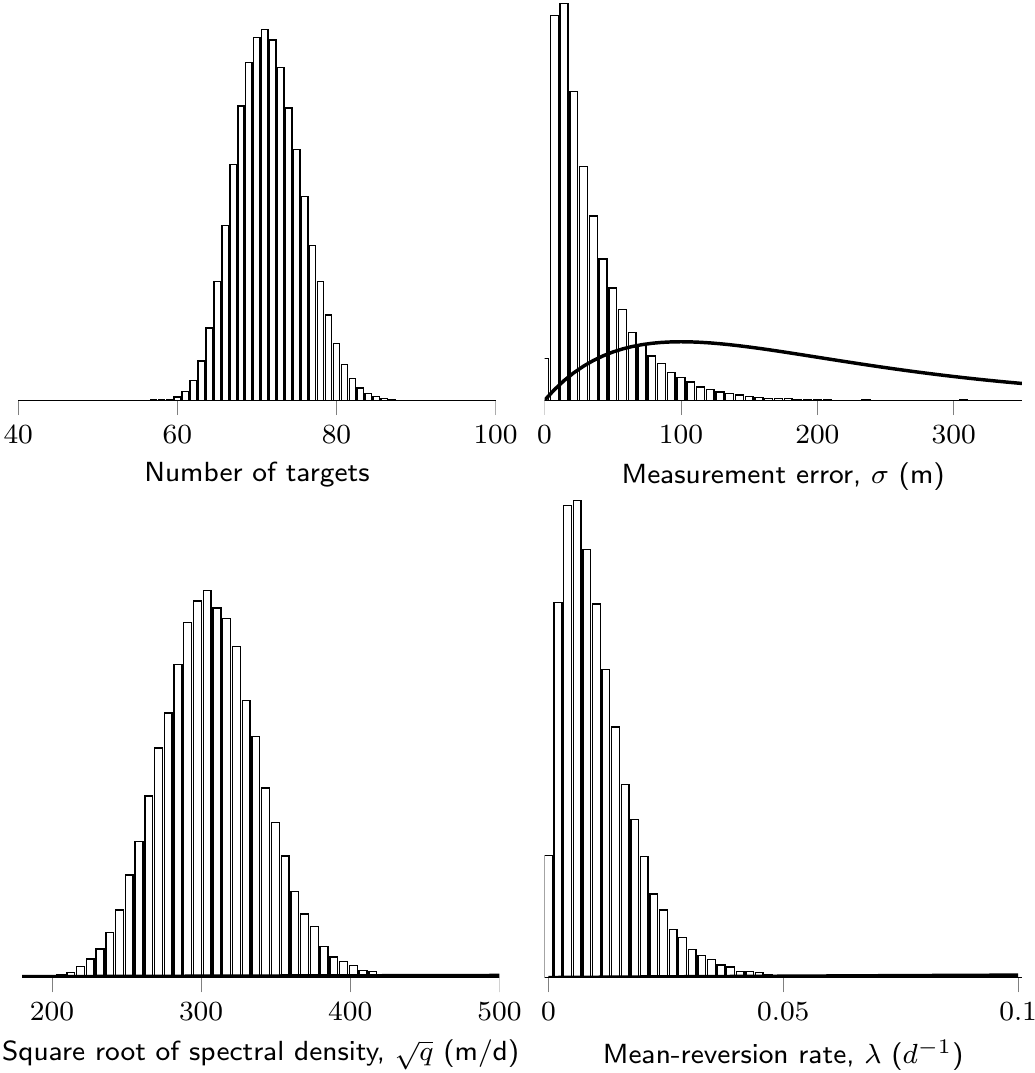}
\caption{Posterior distributions of the number of brown bear families in the Kaakkois-Suomi district in year 2013, and the model parameters. Solid lines denote the corresponding prior densities.}
\label{fig:casehistograms}
\end{figure}

Histograms of the posterior distributions of parameter and number of targets compared to prior densities are shown in Figure~\ref{fig:casehistograms}. Compared to the expert estimates by \emph{Finnish Game and Fisheries Research institute} \citep{expert-estimates}, the model clearly overestimates the number of packs - the expert estimate was $20-22$ while our model predicts about $60-80$ targets. However, this may be due to experts having more information about, for example, which observations are unreliable. Furthermore, it may be that our prior distributions were too noninformative, placing considerable mass on unrealistic parameter values. Indeed, the posterior for the parameter $q$ in the posterior is much smaller than the prior expectation, which naturally explains the high number of targets.

\section{Conclusions and Discussion}
\label{sec:conclusions}
In this paper, we have presented a novel algorithm for parameter estimation in multiple target tracking problems. The algorithm is based on combining the Rao-Blackwellized Monte Carlo data association (RBMCDA) algorithm \citep{sarkka:2007} with particle Markov chain Monte Carlo (PMCMC) methods \citep{andrieu:2010}. We considered two different variations of the algorithm based on the particle marginal Metropolis--Hastings and particle Gibbs algorithms known in the particle MCMC literature. 

In the numeric experiments section, we tested the method with a simulated example and then applied it to a real-data application of estimating the brown bear population in Finland. With the simulated data, we also compared the convergence of the distribution of targets with different variations of our algorithm. 

This research could be continued in several directions. To speed up computations, one could combine gating techniques with RBMCDA \citep{wang:2014}. It may be possible to derive an upper bound for the measurement likelihoods such that early rejection \citep{solonen:2012} could be applied in RBMCDA--PMMH. That is, computational speedup would be obtained by sometimes deducing during a RBMCDA step that a proposal will be rejected, without processing through all measurements. Besides sampling the parameters of the dynamic and measurement models, one could sample the data association priors as well as the initial densities. \citet{sarkka:2007} showed that the RBMCDA algorithm can be easily extended to non-linear models by using an approximative filter, such as the EKF or the UKF \citep{sarkka:2013}. This extension  could as well be combined with PMCMC. Rao-Blackwellized particle smoothing \citep{sarkka:2012,lindsten:2013} could be used to obtain smoothing distributions of the target states. The model could be extended to allow separate parameters for each target. With unknown number of targets, this would require reversible jump MCMC \citep{green:1995,punskaya:2002} or similar techniques. Allowing interaction among target states would enable group tracking \citep[see][and 
references therein]{mihaylova:2014}. Besides particle MCMC, one could investigate other methods combining particle filters with inference on static parameters, such as SMC$^2$ \citep{chopin:2013} and particle learning \citep{carvalho:2010}, in the RBMCDA context.

\section*{Acknowledgments}
This work was supported by grants from the Academy of Finland (266940, 273475). We acknowledge the computational resources provided by the Aalto Science-IT project. We thank Samuli Heikkinen and Mika Kurkilahti from the \emph{Finnish Game and Fisheries Research institute} for providing the real data and helpful discussions. We thank Arno Solin, Joona Karjalainen, and three anonymous reviewers for helpful comments on the manuscript. 

\section*{References}
\bibliographystyle{elsarticle-harv}
\bibliography{sources}

\begin{thebibliography}{67}
\expandafter\ifx\csname natexlab\endcsname\relax\def\natexlab#1{#1}\fi
\expandafter\ifx\csname url\endcsname\relax
  \def\url#1{\texttt{#1}}\fi
\expandafter\ifx\csname urlprefix\endcsname\relax\def\urlprefix{URL }\fi

\bibitem[{Abbas(2011)}]{abbas:2011}
Abbas, M., 2011. Statistical estimation of wild animal population in {F}inland:
  a multiple target tracking approach. Master's thesis, School of Science,
  Aalto University, Finland.

\bibitem[{Akashi and Kumamoto(1977)}]{akashi:1977}
Akashi, H., Kumamoto, H., 1977. Random sampling approach to state estimation in
  switching environments. Automatica 13~(4), 429--434.

\bibitem[{Andrieu et~al.(2010)Andrieu, Doucet, and Holenstein}]{andrieu:2010}
Andrieu, C., Doucet, A., Holenstein, R., 2010. Particle {Markov} chain {Monte
  Carlo} methods. Journal of the Royal Statistical Society: Series~B
  (Statistical Methodology) 72~(3), 269--342.

\bibitem[{Andrieu et~al.(2004)Andrieu, Doucet, Singh, and Tadic}]{andrieu:2004}
Andrieu, C., Doucet, A., Singh, S., Tadic, V., March 2004. Particle methods for
  change detection, system identification, and control. Proceedings of the IEEE
  92~(3), 423--438.

\bibitem[{Bar-Shalom et~al.(2001)Bar-Shalom, Li, Li, and
  Kirubarajan}]{bar-shalom:2001}
Bar-Shalom, Y., Li, X., Li, X., Kirubarajan, T., 2001. {Estimation with
  Applications to Tracking and Navigation}. Wiley-Interscience.

\bibitem[{Blackman and Popoli(1999)}]{blackman:1999}
Blackman, S., Popoli, R., 1999. {Design and Analysis of Modern Tracking
  Systems}. Artech House Norwood, MA.

\bibitem[{Blackman(2004)}]{blackman:2004}
Blackman, S.~S., 2004. Multiple hypothesis tracking for multiple target
  tracking. Aerospace and Electronic Systems Magazine, IEEE 19~(1), 5--18.

\bibitem[{Bugallo et~al.(2007)Bugallo, Lu, and Djuric}]{bugallo:2007}
Bugallo, M.~F., Lu, T., Djuric, P.~M., 2007. Target tracking by multiple
  particle filtering. In: Aerospace Conference, 2007 IEEE. IEEE, pp. 1--7.

\bibitem[{Carvalho et~al.(2010)Carvalho, Johannes, Lopes, and
  Polson}]{carvalho:2010}
Carvalho, C., Johannes, M.~S., Lopes, H.~F., Polson, N., 2010. Particle
  learning and smoothing. Statistical Science 25~(1), 88--106.

\bibitem[{Challa et~al.(2011)Challa, Morelande, Musicki, and
  Evans}]{challa:2011}
Challa, S., Morelande, M., Musicki, D., Evans, R., 2011. Fundamentals of object
  tracking. Cambridge University Press.

\bibitem[{Chen and Liu(2000)}]{chen:2000}
Chen, R., Liu, J.~S., 2000. Mixture {K}alman filters. Journal of the Royal
  Statistical Society: Series B (Statistical Methodology) 62~(3), 493--508.

\bibitem[{Chopin(2010)}]{chopin:2010}
Chopin, N., 2010. Discussion to '{Particle {Markov} chain {Monte Carlo}
  methods}' by {C. Andrieu, A. Doucet and R. Holenstein}. Journal of the Royal
  Statistical Society: Series~B (Statistical Methodology) 72~(3), 304--305.

\bibitem[{Chopin et~al.(2013)Chopin, Jacob, and Papaspiliopoulos}]{chopin:2013}
Chopin, N., Jacob, P.~E., Papaspiliopoulos, O., 2013. {SMC2}: an efficient
  algorithm for sequential analysis of state space models. Journal of the Royal
  Statistical Society: Series B (Statistical Methodology) 75~(3), 397--426.

\bibitem[{Chung et~al.(2013)Chung, Rabe-Hesketh, Dorie, Gelman, and
  Liu}]{chung:2013}
Chung, Y., Rabe-Hesketh, S., Dorie, V., Gelman, A., Liu, J., 2013. A
  nondegenerate penalized likelihood estimator for variance parameters in
  multilevel models. Psychometrika 78~(4), 685--709.

\bibitem[{Clark et~al.(2007)Clark, Vo, and Vo}]{clark:2007-gaussian}
Clark, D., Vo, B.-T., Vo, B.-N., 2007. Gaussian particle implementations of
  probability hypothesis density filters. In: Aerospace Conference, 2007 IEEE.
  IEEE, pp. 1--11.

\bibitem[{Clark and Bell(2007)}]{clark:2007-multi}
Clark, D.~E., Bell, J., 2007. Multi-target state estimation and track
  continuity for the particle {PHD} filter. Aerospace and Electronic Systems,
  IEEE Transactions on 43~(4), 1441--1453.

\bibitem[{Closas and Bugallo(2012)}]{closas:2012}
Closas, P., Bugallo, M.~F., 2012. Improving accuracy by iterated multiple
  particle filtering. Signal Processing Letters, IEEE 19~(8), 531--534.

\bibitem[{Djuric and Bugallo(2009)}]{djuric:2009}
Djuric, P.~M., Bugallo, M.~F., 2009. Improved target tracking with particle
  filtering. In: Aerospace conference, 2009 IEEE. IEEE, pp. 1--7.

\bibitem[{Doucet et~al.(2000{\natexlab{a}})Doucet, De~Freitas, Murphy, and
  Russell}]{doucet:2000-rao}
Doucet, A., De~Freitas, N., Murphy, K., Russell, S., 2000{\natexlab{a}}.
  Rao-{B}lackwellised particle filtering for dynamic {B}ayesian networks. In:
  Proceedings of the Sixteenth conference on Uncertainty in artificial
  intelligence. Morgan Kaufmann Publishers Inc., pp. 176--183.

\bibitem[{Doucet et~al.(2000{\natexlab{b}})Doucet, Godsill, and
  Andrieu}]{doucet:2000-on}
Doucet, A., Godsill, S., Andrieu, C., 2000{\natexlab{b}}. On sequential {Monte
  Carlo} sampling methods for {B}ayesian filtering. Statistics and computing
  10~(3), 197--208.

\bibitem[{Duckworth(2012)}]{duckworth:2012}
Duckworth, D., 2012. Monte {C}arlo methods for multiple target tracking and
  parameter estimation. {B}erkeley technical report.

\bibitem[{Fearnhead and Clifford(2003)}]{fearnhead:2003}
Fearnhead, P., Clifford, P., 2003. On-line inference for hidden {M}arkov models
  via particle filters. Journal of the Royal Statistical Society: Series B
  (Statistical Methodology) 65~(4), 887--899.

\bibitem[{Fern{\'a}ndez-Villaverde and
  Rubio-Ram{\'\i}rez(2007)}]{fernandez-villaverde:2007}
Fern{\'a}ndez-Villaverde, J., Rubio-Ram{\'\i}rez, J.~F., 2007. Estimating
  macroeconomic models: A likelihood approach. The Review of Economic Studies
  74~(4), 1059--1087.

\bibitem[{FGFRI(2014)}]{expert-estimates}
FGFRI, 2014. Population estimates by the {\emph {f}innish game and fisheries
  research institute} (in {F}innish).
  \url{http://www.rktl.fi/riista/suurpedot/rktln_lausunnot_suurpedoista.html}.

\bibitem[{Fortmann et~al.(1980)Fortmann, Bar-Shalom, and
  Scheffe}]{fortmann:1980}
Fortmann, T.~E., Bar-Shalom, Y., Scheffe, M., 1980. Multi-target tracking using
  joint probabilistic data association. In: Decision and Control including the
  Symposium on Adaptive Processes, 1980 19th IEEE Conference on. Vol.~19. IEEE,
  pp. 807--812.

\bibitem[{Gelman et~al.(2013)Gelman, Carlin, Stern, Dunson, Vehtari, and
  Rubin}]{gelman:2013}
Gelman, A., Carlin, J.~B., Stern, H.~S., Dunson, D.~B., Vehtari, A., Rubin,
  D.~B., 2013. Bayesian Data Analysis, 3rd Edition. Chapman \& Hall/CRC.

\bibitem[{Green(1995)}]{green:1995}
Green, P.~J., 1995. Reversible jump {M}arkov chain {M}onte {C}arlo computation
  and {B}ayesian model determination. Biometrika 82~(4), 711--732.

\bibitem[{Haario et~al.(2001)Haario, Saksman, and Tamminen}]{haario:2001}
Haario, H., Saksman, E., Tamminen, J., 2001. An adaptive {M}etropolis
  algorithm. Bernoulli, 223--242.

\bibitem[{Hwang et~al.(2004)Hwang, Balakrishnan, Roy, and Tomlin}]{hwang:2004}
Hwang, I., Balakrishnan, H., Roy, K., Tomlin, C., 2004. Multiple-target
  tracking and identity management in clutter, with application to aircraft
  tracking. In: American Control Conference, 2004. Proceedings of the 2004.
  Vol.~4. IEEE, pp. 3422--3428.

\bibitem[{Jones et~al.(2010)Jones, Parslow, and Murray}]{jones:2010}
Jones, E., Parslow, J., Murray, L., 2010. A {B}ayesian approach to state and
  parameter estimation in a phytoplankton-zooplankton model. Australian
  Meteorological and Oceanographic Journal 59, 7--16.

\bibitem[{Kalman(1960)}]{kalman:1960}
Kalman, R.~E., 1960. A new approach to linear filtering and prediction
  problems. Journal of Fluids Engineering 82~(1), 35--45.

\bibitem[{Kirubarajan and Bar-Shalom(2005)}]{kirubarajan:2005}
Kirubarajan, T., Bar-Shalom, Y., 2005. {Probabilistic data association
  techniques for target tracking in clutter}. Proceedings of the {IEEE} 92~(3),
  536--557.

\bibitem[{Kojola(2007)}]{kojola:2007}
Kojola, I., 2007. Biology of the bear and the current status of the bear
  population. In: Management Plan for the Bear Population in Finland.
  Publications of Ministry of Agriculture and Forestry, Finland, pp. 10--17.

\bibitem[{Lindsten et~al.(2013)Lindsten, Bunch, Godsill, and
  Schon}]{lindsten:2013}
Lindsten, F., Bunch, P., Godsill, S.~J., Schon, T.~B., 2013.
  {R}ao-{B}lackwellized particle smoothers for mixed linear/nonlinear
  state-space models. In: Acoustics, Speech and Signal Processing ({ICASSP}),
  2013 {IEEE} International Conference on. IEEE, pp. 6288--6292.

\bibitem[{Liu and Chen(1995)}]{liu:1995}
Liu, J.~S., Chen, R., 1995. Blind deconvolution via sequential imputations.
  Journal of the American Statistical Association 90~(430), 567--576.

\bibitem[{Mahler(2003)}]{mahler:2003}
Mahler, R. P.~S., 2003. Multitarget {B}ayes filtering via first-order
  multitarget moments. {IEEE} Transactions on Aerospace and Electronic Systems
  39~(4), 1152--1178.

\bibitem[{Mahler(2007{\natexlab{a}})}]{mahler:2007}
Mahler, R. P.~S., 2007{\natexlab{a}}. {PHD} filters of higher order in target
  number. {IEEE} Transactions on Aerospace and Electronic Systems 43~(4),
  1523--1543.

\bibitem[{Mahler(2007{\natexlab{b}})}]{mahlerbook:2007}
Mahler, R. P.~S., 2007{\natexlab{b}}. Statistical Multisource-Multitarget
  Information Fusion. Artech House.

\bibitem[{Martino et~al.(2015)Martino, Leisen, and Corander}]{Martino:2015}
Martino, L., Leisen, F., Corander, J., 2015. On multiple try schemes and the
  particle {M}etropolis--{H}astings algorithm, viXra preprint:
  http://www.rxiv.org/pdf/1409.0051v2.pdf.

\bibitem[{Mestre and Fitzgerald(2013)}]{mestre:2013}
Mestre, M.~R., Fitzgerald, W.~J., 2013. Multi-target tracking applied to
  evolutionary clustering. In: Acoustics, Speech and Signal Processing
  ({ICASSP}), 2013 {IEEE} International Conference on. IEEE, pp. 3173--3177.

\bibitem[{Mihaylova et~al.(2014)Mihaylova, Carmi, Septier, Gning, Pang, and
  Godsill}]{mihaylova:2014}
Mihaylova, L., Carmi, A.~Y., Septier, F., Gning, A., Pang, S.~K., Godsill, S.,
  2014. Overview of {B}ayesian sequential {M}onte {C}arlo methods for group and
  extended object tracking. Digital Signal Processing 25, 1--16.

\bibitem[{Minvielle et~al.(2014)Minvielle, Todeschini, Caron, and
  Del~Moral}]{minvielle:2014}
Minvielle, P., Todeschini, A., Caron, F., Del~Moral, P., 2014. Particle {MCMC}
  for {B}ayesian microwave control. ar{X}iv preprint ar{X}iv:1405.2673.

\bibitem[{Nevat et~al.(2011)Nevat, Peters, and Yuan}]{nevat:2011}
Nevat, I., Peters, G.~W., Yuan, J., 2011. Channel tracking in relay systems via
  particle {MCMC}. In: Vehicular Technology Conference ({VTC} Fall), 2011
  {IEEE}. IEEE, pp. 1--5.

\bibitem[{Orton and Fitzgerald(2002)}]{orton:2002}
Orton, M., Fitzgerald, W., 2002. A {B}ayesian approach to tracking multiple
  targets using sensor arrays and particle filters. Signal Processing, IEEE
  Transactions on 50~(2), 216--223.

\bibitem[{Peters and Cornebise(2010)}]{peters:2010}
Peters, G., Cornebise, J., 2010. Discussion to '{Particle {Markov} chain {Monte
  Carlo} methods}' by {C. Andrieu, A. Doucet and R. Holenstein}. Journal of the
  Royal Statistical Society: Series~B (Statistical Methodology) 72~(3),
  304--305.

\bibitem[{Peters et~al.(2013)Peters, Briers, Shevchenko, and
  Doucet}]{peters:2013}
Peters, G.~W., Briers, M., Shevchenko, P., Doucet, A., 2013. Calibration and
  filtering for multi factor commodity models with seasonality: incorporating
  panel data from futures contracts. Methodology and Computing in Applied
  Probability 15~(4), 841--874.

\bibitem[{Petetin et~al.(2014)Petetin, Morelande, and
  Desbouvries}]{petetin:2014}
Petetin, Y., Morelande, M., Desbouvries, F., 2014. Marginalized particle {PHD}
  filters for multiple object {B}ayesian filtering. Aerospace and Electronic
  Systems, IEEE Transactions on 50~(2), 1182--1196.

\bibitem[{Punskaya et~al.(2002)Punskaya, Andrieu, Doucet, and
  Fitzgerald}]{punskaya:2002}
Punskaya, E., Andrieu, C., Doucet, A., Fitzgerald, W.~J., 2002. Bayesian curve
  fitting using {MCMC} with applications to signal segmentation. Signal
  Processing, {IEEE} Transactions on 50~(3), 747--758.

\bibitem[{Rao and Satyanarayana(2013)}]{rao:2013}
Rao, G.~M., Satyanarayana, C., 2013. Visual object target tracking using
  particle filter: A survey. International Journal of Image, Graphics and
  Signal Processing 5~(6), 1250.

\bibitem[{Ravindra et~al.(2012)Ravindra, Svensson, Hammarstrand, and
  Morelande}]{ravindra:2012}
Ravindra, V.~C., Svensson, L., Hammarstrand, L., Morelande, M., 2012. A
  cardinality preserving multitarget multi-{B}ernoulli {RFS} tracker. In:
  Information Fusion ({FUSION}), 2012 15th International Conference on. IEEE,
  pp. 832--839.

\bibitem[{Reid(1979)}]{reid:1979}
Reid, D.~B., 1979. An algorithm for tracking multiple targets. Automatic
  Control, {IEEE} Transactions on 24~(6), 843--854.

\bibitem[{S{\"a}rkk{\"a}(2013)}]{sarkka:2013}
S{\"a}rkk{\"a}, S., 2013. Bayesian Filtering and Smoothing. Vol.~3 of Institute
  of Mathematical Statistics Textbooks. Cambridge University Press.

\bibitem[{S{\"a}rkk{\"a} et~al.(2012)S{\"a}rkk{\"a}, Bunch, and
  Godsill}]{sarkka:2012}
S{\"a}rkk{\"a}, S., Bunch, P., Godsill, S., 2012. A backward-simulation based
  {R}ao-{B}lackwellized particle smoother for conditionally linear {G}aussian
  models. In: Proceedings of the 16th IFAC Symposium on System Identification,
  Brussels, Belgium.

\bibitem[{S{\"a}rkk{\"a} et~al.(2007)S{\"a}rkk{\"a}, Vehtari, and
  Lampinen}]{sarkka:2007}
S{\"a}rkk{\"a}, S., Vehtari, A., Lampinen, J., 2007. {{R}ao-{B}lackwellized
  particle filter for multiple target tracking}. Information Fusion 8~(1),
  2--15.

\bibitem[{Schuhmacher et~al.(2008)Schuhmacher, Vo, and Vo}]{schuhmacher:2008}
Schuhmacher, D., Vo, B.-T., Vo, B.-N., 2008. A consistent metric for
  performance evaluation of multi-object filters. Signal Processing, IEEE
  Transactions on 56~(8), 3447--3457.

\bibitem[{Solonen et~al.(2012)Solonen, Ollinaho, Laine, Haario, Tamminen,
  J{\"a}rvinen, et~al.}]{solonen:2012}
Solonen, A., Ollinaho, P., Laine, M., Haario, H., Tamminen, J., J{\"a}rvinen,
  H., et~al., 2012. Efficient {MCMC} for climate model parameter estimation:
  Parallel adaptive chains and early rejection. Bayesian Analysis 7~(3),
  715--736.

\bibitem[{Svensson and Morelande(2014)}]{svensson:2014}
Svensson, L., Morelande, M., 2014. Target tracking based on estimation of sets
  of trajectories. In: Information Fusion ({FUSION}), 2014 17th International
  Conference on. IEEE, pp. 1--8.

\bibitem[{Svensson et~al.(2011)Svensson, Svensson, Guerriero, and
  Willett}]{svensson:2011}
Svensson, L., Svensson, D., Guerriero, M., Willett, P., 2011. Set {JPDA} filter
  for multitarget tracking. Signal Processing, {IEEE} Transactions on 59~(10),
  4677--4691.

\bibitem[{Vanhatalo et~al.(2013)Vanhatalo, Riihim{\"a}ki, Hartikainen,
  Jyl{\"a}nki, Tolvanen, and Vehtari}]{vanhatalo:2013}
Vanhatalo, J., Riihim{\"a}ki, J., Hartikainen, J., Jyl{\"a}nki, P., Tolvanen,
  V., Vehtari, A., 2013. {GP}stuff: {B}ayesian modeling with {G}aussian
  processes. The Journal of Machine Learning Research 14~(1), 1175--1179.

\bibitem[{Vihola(2007)}]{vihola:2007}
Vihola, M., 2007. Rao-{B}lackwellised particle filtering in random set
  multitarget tracking. Aerospace and Electronic Systems, IEEE Transactions on
  43~(2), 689--705.

\bibitem[{Vo and Ma(2006)}]{vo:2006}
Vo, B.-N., Ma, W.-K., 2006. The {G}aussian mixture probability hypothesis
  density filter. Signal Processing, {IEEE} Transactions on 54~(11),
  4091--4104.

\bibitem[{Vo et~al.(2003)Vo, Singh, and Doucet}]{vo:2003}
Vo, B.-N., Singh, S., Doucet, A., 2003. Sequential {M}onte {C}arlo
  implementation of the {PHD} filter for multi-target tracking. In: Proc.
  Int’l Conf. on Information Fusion. pp. 792--799.

\bibitem[{Vo et~al.(2009)Vo, Vo, and Cantoni}]{vo:2009}
Vo, B.-T., Vo, B.-N., Cantoni, A., 2009. The cardinality balanced multi-target
  multi-{B}ernoulli filter and its implementations. Signal Processing, {IEEE}
  Transactions on 57~(2), 409--423.

\bibitem[{Vu et~al.(2014)Vu, Vo, and Evans}]{vu:2014}
Vu, T., Vo, B.-N., Evans, R., 2014. A particle marginal {M}etropolis-{H}astings
  multi-target tracker. Signal processing, {IEEE} Transactions on 62~(15),
  3953--3964.

\bibitem[{Wang and Zhang(2014)}]{wang:2014}
Wang, Y., Zhang, P., 2014. Gating techniques for {Rao-Blackwellized Monte Carlo
  Data Association} filter. The Scientific World Journal 2014.

\bibitem[{Whiteley et~al.(2010)Whiteley, Andrieu, and Doucet}]{whiteley:2010}
Whiteley, N., Andrieu, C., Doucet, A., 2010. Efficient {B}ayesian inference for
  switching state-space models using discrete particle {M}arkov chain {M}onte
  {C}arlo methods. arXiv preprint arXiv:1011.2437.

\bibitem[{Yi et~al.(2013)Yi, Morelande, Kong, and Yang}]{yi:2013}
Yi, W., Morelande, M.~R., Kong, L., Yang, J., 2013. A computationally efficient
  particle filter for multitarget tracking using an independence approximation.
  Signal Processing, {IEEE} Transactions on 61~(4), 843--856.

\end{thebibliography}
\pagebreak
\appendix
\section{Algorithms}
\label{app:algs}

In this section, we present the algorithms discussed in the paper in pseudocode. 
\tiny
\begin{algorithm}
 \tiny
 \begin{algorithmic}
 \Require{State mean $\vm$ and state covariance $\MP$ after step $k-1$. Time step $k$.}
 \Ensure{Predicted state mean $\vm^{-}$ and state covariance $\MP^{-}$ at time step $k$ without conditioning on measurements.}
   \Function{Predict}{$\vm,\MP,k$}
   \State $\vm^{-} \gets \MA_{k-1}\vm$
   \State $\MP^{-} \gets \MA_{k-1}\,\MP\,\MA_{k-1}^{\T} + \MQ_{k-1}$
   \EndFunction
   \end{algorithmic}
   \caption{The Kalman filter prediction step.}
  \label{alg:predict}
\end{algorithm}

\begin{algorithm}
 \tiny
 \begin{algorithmic}
 \Require{Measurements $\vy_{1:M}$. Model parameters $\vtheta$. Number of particles $N$.}
 \Ensure{Samples of the data association histories and corresponding weights: $\left(c^{(1:N)}_{1:T},w^{(1:N)}\right)$, likelihood approximation $\hat{p}(\vy_{1:T} \mid \vtheta)$.}
 \Function{RBMCDA}{$\vy_{1:M}, \vtheta, N$}
 \For{$i=1,\ldots,N$} \Comment{Initialize the particles}
    \State $w^{(i)} = 1/N$
    \State $T^{(i)}_0 = 0$
 \EndFor
 \State $\ve^{(1:N)}_0 \gets \emptyset$ \Comment{No targets exist initially}
 \State $\hat{p}(\vy_{1:0} \mid \vtheta) \gets 1$ \Comment{Likelihood approximation}
 \For{$k=1,\ldots,M$}
   \For{$i = 1,\ldots,N$}
     \For{$j = 1,\ldots, T_{k-1}^{(i)}$}
       \State $\vm^{(i)}_{k,j},\MP^{(i)}_{k,j} \gets $ Predict($\vm^{(i)}_{k-1,j},\MP^{(i)}_{k-1,j},k$)
     \EndFor
     \State $\vm^\ast,\MP^\ast,\pi \gets$
     \State $\quad$ EvalImpDist($\vm^{(i)}_{1:T_{k-1},k},\MP^{(i)}_{1:T_{k-1},k},\vy,c^{(i)}_{1:k-1},\ve^{(i)}_{k-1})$ %
     \State $v^{(i)}_k \gets w^{(i)}_{k-1}\times \sum \pi_j$
     \State $\forall j \in \{1,\ldots,T^{(i)}_{k-1}+1\}:~\pi_j \gets \frac{\pi_j}{\sum \pi_j}$
     \State Draw $l$ with probabilities $\left(\pi_1,\ldots,\pi_{T^{(i)}_{k-1}+1}\right)$
     \State $\ve^{(i)}_k \gets \ve^{(i)}_{k-1}$     
     \If{$l\neq 0$}
      \State $\vm^{(i)}_{k,l},\MP^{(i)}_{k,l} \gets \vm^{\ast}_l,\MP^\ast_l$
     \EndIf
     \If{$l = T^{(i)}_{k-1} + 1$}
      \State $T^{(i)}_k \gets T^{(i)}_{k-1} + 1$
      \State $\ve^{(i)}_k(l) \gets 1$
     \Else
      \State $T^{(i)}_k \gets T^{(i)}_{k-1}$
     \EndIf
     \Comment{Remove targets:}
     \For{$m \in \{1,\ldots,T^{(i)}_k\}$}
      \If{Time since last observation associated to $m$ in particle $i$ $>$ threshold}
        \State $\ve^{(i)}_k(m) \gets 0$
        \State $T^{(i)}_k \gets T^{(i)}_{k} - 1$
      \EndIf
     \EndFor
   \EndFor
   \State $\hat{p}(\vy_k \mid \vy_{1:{k-1}}, \vtheta) \gets \sum_{i=1}^N v_k^{(i)}w^{(i)}_{k-1}$
   \State $\hat{p}(\vy_{1:k} \mid \vtheta) \gets \hat{p}(\vy_k \mid \vy_{1:{k-1}}, \vtheta)\,\hat{p}(\vy_{1:{k-1}} \mid \vtheta)$    
   \State $\forall i:~w^{(i)}_k \gets \frac{v^{(i)}_k}{\sum_{j=1}^N v^{(j)}_k}$
   \State Resample.
 \EndFor
 \EndFunction
 \end{algorithmic}
 \caption{The Rao-Blackwellized Monte Carlo data association algorithm.}
 \label{alg:rbmcda}
\end{algorithm}

\begin{algorithm}
 \tiny
 \begin{algorithmic}
 \Require{Predicted state mean $\vm^{-}$, state covariance $\MP^{-}$, measurement $\vy$ and time step $k$.}
 \Ensure{Updated mean $\vm$ and covariance $\MP$ of the state distribution conditional on the measurement $\vy$. Likelihood $\mathrm{lh}$ of the measurement.}
   \Function{Update}{$\vm^{-},\MP^{-},\vy,k$}
   \State $\vv \gets \vy - \MH_k\vm^{-}$
   \State $\MS \gets \MH_k\,\MP^{-}\,\MH_k^{\T} + \MR_k$
   \State $\MK \gets \MP^{-}\MH_k^{\T}\MS^{-1}$
   \State $\mathrm{lh} \gets |2\pi\MS|^{-\frac{1}{2}}\,e^{-\frac{1}{2}\vv^{\T}\,\MS^{-1}\vv}$
   \State $\vm \gets \vm^{-} + \MK\,\vv$
   \State $\MP \gets \MP^{-} - \MK\,\MS\,\MK^{\T}$
   \EndFunction
   \end{algorithmic}
   \caption{Kalman filter update step.}
  \label{alg:update}
\end{algorithm}

\begin{algorithm}
 \tiny
 \begin{algorithmic}
 \Require{Predicted target state distribution moments $\vm_{1:T},\MP_{1:T}$, measurement $\vy$, number of targets $T$, association history $c_{1:k-1}$, visibility indicator $\ve_{k-1}$. Implicitly: time step $k$, time- and model specific Update function performing the Kalman filter update step and evaluating measurement likelihood.}
 \Ensure{Unnormalized optimal importance distribution $(\pi_1,\ldots,\pi_{T+1})$, target state distribution moments $\vm_{1:{T+1}}^\ast,\MP_{1:{T+1}}^\ast$ conditional to associating the measurement to each particular target.
Optionally (cf. Alg.~\ref{alg:crbmcda}) returns also the measurement likelihoods $\mathrm{lh}_{1:T+1}$.}
 \Function{EvalImpDist}{$\vm_{1:T},\MP_{1:T},\vy,c_{1:k-1},\ve_{k-1}$}
     \State $\mathrm{lh}_0 \gets p(\vy \mid c_k = 0)$
     \State $\pi_0 \gets p(c_k = 0 \mid c_{1:k-1},\ve_{k-1})$      
     \For{$j = 1,\ldots, T$}
       \If{$\ve_{k-1}(j) = 1$}
        \State $(\vm^{\ast}_j,\MP^{\ast}_j,\mathrm{lh}_j) \gets $ Update($\vm_{j},\MP_j,\vy,k$)
        \State $\pi_j \gets \mathrm{lh}_j\times p(c_k = j \mid c_{1:k-1},\ve_{k-1})$ 
       \EndIf 
     \EndFor
     \State $\left(\vm^{\ast}_{T+1},\MP^{\ast}_{T+1},\mathrm{lh}_{T+1}\right) \gets $ Update($\vm_0,\MP_0,\vy,k$) 
     \State $\pi_{T+1} \gets \mathrm{lh}_{T+1}\times p(c_k = T+1 \mid c_{1:k-1})$
 \EndFunction
 \end{algorithmic}
 \caption{Algorithm for evaluating the unnormalized optimal importance distribution and updated target states conditional on associations.}
 \label{alg:evalimpdistr}
\end{algorithm}

\begin{algorithm}
\tiny
\begin{algorithmic}
\Require{Measurements $\vy_{1:T}$, initial parameters $\vtheta^0$. Sample size $I$. Number of particles used in RBMCDA ($N$). Covariance adaptation period $i_1,i_2$ Implicitly: dimension of parameters $d$, model-specific functions Update,Predict used in RBMCDA.}
\Ensure{Samples from the posterior distribution of parameters, $\vtheta^1,\vtheta^2,\ldots,\vtheta^i$. Weighted samples from the (marginal) posterior distribution of data association histories, $\forall i\in\{0,\ldots,I\}:~u_i,w_T^{(1:N),i}, c^{(1:N),i}_{1:T}$ where the total weight of data association history $c^{(j),i}_{1:T}$ is $u_i\,w^{(j)}_i$. } 
\Function{PMMH}{$\vy_{1:T},\vtheta^0$}  
  \State $\Sigma \gets \Sigma_0$ \Comment{Initialize proposal covariance}
  \State $\left(c^{(1:N)}_{1:T},w^{(1:N),0}_{T}\right),\hat{p}^0 \gets$ RBMCDA($\vy_{1:T},\vtheta^0$) 
  \State lastaccept $\gets 0$ \Comment{Used to update the weights $u$}
  \For{$i = 1,2,\ldots,I$}
    \If{$i_1\leq i \leq i_2$}
    \State $\Sigma \gets (\frac{2.4}{d})^2\mathrm{Cov}(\theta^0,\ldots,\theta^i) + \epsilon\,I_d$
    \EndIf
    \State Draw $\vtheta^\ast \sim \N(\vtheta^\ast \mid \vtheta^{i-1}, \Sigma)$ 
    \State $\left(c^{(1:N),i}_{1:T},w^{(1:N),i}_{T}\right),\hat{p}^\ast \gets$ RBMCDA($\vy_{1:T},\vtheta^\ast,N$) 
    \State $\alpha \gets \min\left(1, \frac{\hat{p}^\ast p(\vtheta^\ast)}{\hat{p}^{i-1}p(\vtheta^{i-1})}\right)$
    \State $u_i \gets \alpha$, $u_\textrm{lastaccept} \gets  u_\textrm{lastaccept} + (1-\alpha)$
    \State Draw $Z \sim \mathrm{U}(0,1)$
    \If{$Z < \alpha$} 
     \State $\vtheta^i,\hat{p} \gets \vtheta^\ast,\vx^\ast_{0:T},\hat{p}^\ast$
     \State lastaccept $\gets i$
    \Else
     \State $\vtheta^i,\hat{p}^i \gets \vtheta^{i-1},\hat{p}^{i-1}$
    \EndIf
  \EndFor
\EndFunction
\end{algorithmic}
 \caption{The particle marginal Metropolis--Hastings algorithm with RBMCDA.}
 \label{alg:rbmcda_pmmh}
\end{algorithm}

\begin{algorithm}
 \tiny
 \begin{algorithmic}
 \Require{Measurements $\vy_{1:M}$. Model parameters $\vtheta$. Number of particles $N$. Fixed data association history $c^{(1)}_{1:T}$}
 \Ensure{Samples of the data association histories and corresponding weights: $\left(c^{(1:N)}_{1:T},w_T^{(1:N)}\right)$.
Conditional likelihood for each data association history: $p_{i\in\{1,\ldots,I\}}(\vy \mid \vtheta,c_{1:T}^{(i)})$}
 \Function{CRBMCDA}{$\vy_{1:M},\vtheta,N,c_{1:T}^{(1)}$}
 \For{$i=1,\ldots,N$} \Comment{Initialize the particles}
    \State $w^{(i)} \gets 1/N$
    \State $T^{(i)}_0 \gets 0$
 \EndFor
 \For{$k=1,\ldots,M$}
   \For{$i = 1,\ldots,N$}
     \For{$j = 1,\ldots, T_{k-1}^{(i)}$}
       \State $\vm^{(i)}_{k,j},\MP^{(i)}_{k,j} \gets $ Predict($\vm^{(i)}_{k-1,j},\MP^{(i)}_{k-1,j},k$)
     \EndFor
     \State $\left(\vm^\ast,\MP^\ast,\pi, \mathrm{lh}\right)_{1:T^{(i)}_{k-1}+1} \gets$
     \State $\qquad$ EvalImpDist($\vm^{(i)}_{1:T_{k-1},k},\MP^{(i)}_{1:T_{k-1},k},\vy,c^{(i)}_{1:k-1},k)$ %
     \State $v^{(i)}_k \gets w^{(i)}_{k-1}\times \sum \pi_j$
     \State $\forall j \in \{1,\ldots,T^{(i)}_{k-1}+1\}:~\pi_j \gets \frac{\pi_j}{\sum \pi_j}$
     \If{$i>1$}
      \State Draw $l$ with probabilities $\pi_{1:T^{(i)}_{k-1}+1}$%
     \Else
      \State $l \gets c^{(1)}_k$
     \EndIf
     \State $\ve_k \gets \ve_{k-1}$     
     \If{$l\neq 0$}
      \State $\vm^{(i)}_{k,l},\MP^{(i)}_{k,l} \gets \vm^{\ast}_l,\MP^\ast_l$
     \EndIf
     \If{$l = T^{(i)}_{k-1} + 1$}
      \State $T^{(i)}_k \gets T^{(i)}_{k-1} + 1$
      \State $\ve^{(i)}_k(l) \gets 1$
     \Else
       \State $T^{(i)}_k \gets T^{(i)}_{k-1}$
     \EndIf    
     \Comment{Remove targets:}
     \For{$m \in \{1,\ldots,T^{(i)}_k\}$}
      \If{Time since last observation associated to $m$ in particle $i$ $>$ threshold}
        \State $\ve^{(i)}_k(m) \gets 0$
        \State $T^{(i)}_k \gets T^{(i)}_k - 1$
      \EndIf
     \EndFor    
     \State $\vm^{(i)}_{k,l},\MP^{(i)}_{k,l} \gets \vm^{\ast}_l,\MP^\ast_l$
   \EndFor
   \State $\forall i:~w^{(i)}_k \gets \frac{v^{(i)}_k}{\sum_{j=1}^N v^{(j)}_k}$
   \State Resample -- first particle is not changed.

 \EndFor
 \EndFunction
 \end{algorithmic}
 \caption{The conditional Rao-Blackwellized Monte Carlo data association algorithm.}
 \label{alg:crbmcda}
\end{algorithm}

\begin{algorithm}
 \tiny
 \begin{algorithmic}
 \Require{Measurements $\vy_{1:T}$. Model parameters $\vtheta$. Data association history $c_{1:T}$. }
 \Ensure{Likelihood $p(\vy_{1:T} \mid \vtheta, c_{1:T})$.}
 \Function{EvaluateLH}{$\vy_{1:T},~\vtheta,~c_{1:T}$}
   \State $N \gets 0$ \Comment{Targets seen so far}
   \State $p(\vy_{1:0} \mid \vtheta, c_{1:0}) \gets 1$ \Comment{Initialize likelihood}
   \For{$k=1,\ldots,T$}
     \For{$i=1,\ldots,N$}
        \State $\vm_i,\MP_i \gets $ Predict($\vm_i,\MP_i,k,\vtheta$)
     \EndFor
     \If{$c_k = N+1$} \Comment{New target}
       \State $\vm_{N+1},\MP_{N+1} \gets \vm_0,\MP_0$
       \State $N \gets N + 1$
     \EndIf
     \State $\vm_{c_k},\MP_{c_k},\mathrm{lh} \gets $Update($\vm_{c_k},\MP_{c_k},\vy_k,k,\vtheta$)
     \State $p(\vy_{1:k} \mid \vtheta, c_{1:k}) \gets \mathrm{lh} \times p(\vy_{1:k-1} \mid \vtheta, c_{1:k-1})$
   \EndFor 
 \EndFunction
 \end{algorithmic}
 \caption{Evaluating the likelihood conditional on a given data association history.}
 \label{alg:likelihood}
\end{algorithm}

\begin{algorithm}
\tiny
\begin{algorithmic}
\Require{Measurements $\vy_{1:T}$, initial parameters $\vtheta^0$. Sample size $I$. Number of particles used in RBMCDA ($N$). Covariance adaptation period $i_1,i_2$ Implicitly: dimension of parameters $d$, model-specific functions Update and Predict used in RBMCDA.}
\Ensure{Samples from the posterior distribution of parameters, $\vtheta^1,\vtheta^2,\ldots,\vtheta^i$. Weighted samples from the (marginal) posterior distribution of data association histories, $\forall i\in\{0,\ldots,I\}:w_T^{(1:N),i}, c^{(1:N),i}_{1:T}$. } 
\Function{PGibbs}{$\vy_{1:T},\vtheta^0$}
  \State $c^{(1:N),0}_{1:T},w^{(1:N),0}_T \gets $ RBMCDA($\vy_{1:T},\vtheta^0,N$) 
  \State Draw $l \in \{1,\ldots,N\}$ with probabilities $w^{(1:N),0}_T$
  \State $\bar{c}_{1:T} \gets c^{(l),0}_{1:T}$
  \State $p(\vy \mid \vtheta^0 , \bar{c}_{1:T}) \gets$ EvaluateLH($\vy_{1:T},\vtheta^0,\bar{c}_{1:T}$)
  \State $\Sigma \gets \Sigma_0$ \Comment{Initialize proposal covariance}
  \For{$i = 1,2,\ldots,I$}
    \If{$i_1\leq i \leq i_2$}
    \State $\Sigma \gets (\frac{2.4}{d})^2\mathrm{Cov}(\theta^0,\ldots,\theta^i) + \epsilon\,I_d$
    \EndIf
    \State Draw $\vtheta^\ast \sim \N(\vtheta^\ast \mid \vtheta^{i-1}, \Sigma)$
    \State $p(\vy \mid \vtheta^\ast, \bar{c}_{1:T}) \gets$ EvaluateLH($\vy_{1:T},\vtheta^\ast,\bar{c}_{1:T})$ 
    \State Draw $Z \sim U(0,1)$
    \If{$Z < \frac{p(\vtheta^\ast)p(\vy \mid \vtheta^\ast, \bar{c}_{1:T})}{p(\vtheta^{i-1})p(\vy \mid \vtheta^{i-1}, \bar{c}_{1:T})}$}
      \State $\vtheta^i \gets \vtheta^\ast$
    \Else
      \State $\vtheta^i \gets \vtheta^{i-1}$
    \EndIf
    \State $c^{(1:N),i}_{1:T}, w^{(1:N),i}_T, p_{j\in\{1,\ldots,N\}}(\vy \mid \vtheta^i,c^{(j),i}_{1:T}) \gets $ 
    \State $\qquad$ CRBMCDA($\vy_{1:T},\vtheta^i,N,\bar{c}_{1:T}$)
    \State Draw $l \in \{1,\ldots,N\}$ with probabilities $w^{(1:N),i}_T$
    \State $\bar{c}_{1:T}, p(\vy \mid \vtheta^i, \bar{c}_{1:T}) \gets c^{(l)}_{1:T}, p(\vy \mid \vtheta^i,c^{(l)i}_{1:T})$
  \EndFor
\EndFunction
\end{algorithmic}
\caption{The particle Gibbs algorithm with RBMCDA.}
\label{alg:rbmcda_pgibbs}
\end{algorithm}

\end{document}